\documentclass[aps,prd,superscriptaddress,floatfix,nofootinbib,preprintnumbers,eqsecnum,twocolumn,dvipsnames]{revtex4-2}
\counterwithout{equation}{section}
\usepackage{amsmath,float,graphicx,placeins,amssymb,multirow,pgfplots,pgfplotstable}
\usepackage{graphicx,pstricks,listings,stackengine}
\usepackage{lipsum}
\usepackage{empheq}
\usepackage{tikz}
\usepackage{comment}
\usepackage{enumerate,slashed,cancel,comment,color,bbm,graphicx,rotating,placeins,mathtools,multirow,hhline}
\usepackage[normalem]{ulem}
\usepackage{array}
\usepackage{mathrsfs}
\usepackage{hyperref}
\usepackage{color}
 \hypersetup
 {
   colorlinks,
   linkcolor={blue!80!black},
   citecolor={blue!70!black},
   urlcolor={blue!70!black}
 }

\usepackage{lipsum}
\usepackage{diagbox}
\usetikzlibrary{shapes.geometric,arrows}
\graphicspath{}

\def\beq{\begin{equation}}
\def\eeq{\end{equation}}
\def\beqn{\begin{eqnarray}}
\def\eeqn{\end{eqnarray}}

\def\be{\begin{equation}}
\def\ee{\end{equation}}
\def\bea{\begin{eqnarray}}
\def\eea{\end{eqnarray}}

\begin{document}

\title{High-Frequency Thermal Graviton Remnant from the End of Inflation}

\author{Chen-Hao Wu}
 \email[Email(Corresponding Author):]{chenhao\_wu@nuaa.edu.cn}
 \affiliation{College of Physics, Nanjing University of Aeronautics and Astronautics,\\ Nanjing, 210016, China}
 \affiliation{Center for the Cross-disciplinary Research of Space Science and Quantum-technologies (CROSS-Q), NUAA, Nanjing, 210016, China}
\author{Xiao Liang}
 \email[Email:]{xliang@nuaa.edu.cn}
 \affiliation{College of Physics, Nanjing University of Aeronautics and Astronautics,\\ Nanjing, 210016, China}
\author{Ya-Peng Hu}
\email[Email:]{huyp@nuaa.edu.cn}
 \affiliation{College of Physics, Nanjing University of Aeronautics and Astronautics,\\ Nanjing, 210016, China}
 \affiliation{Center for the Cross-disciplinary Research of Space Science and Quantum-technologies (CROSS-Q), NUAA, Nanjing, 210016, China}
  \affiliation{MIIT Key Laboratory of Aerospace Information Materials and Physics,  Nanjing University of Aeronautics and Astronautics, Nanjing, 210016, China}

\begin{abstract} 
The standard inflationary theory focuses on the freezing of super-horizon fluctuations, which generate a scale-invariant spectrum, while the sub-horizon modes are expected to remain in thermal equilibrium based on the perspective of quantum thermodynamics of the de Sitter universe from [Alicki et al. Phys.Rev.D 108 (2023) 12, 123530]. Given this framework, we investigate a concrete observable consequence of the graviton remnant originating from this thermal horizon radiation released at the end of inflation. Unlike the stochastic background from super-horizon fluctuations, this signal might represent a snapshot of the thermal dS state, which subsequently decouples and undergoes cosmological redshift. We present a semi-analytical approximation prediction for this relic background, typically peaking in near MHz band, with characteristic energy density of $\log_{10}(\Omega_{\rm G} h^2) \sim \mathcal{O}(-18)$. These signals occupy a High-Frequency band, offering a potential novel probe of the reheating temperature and the thermal history of the early universe.
\end{abstract}

\maketitle


\def\beq{\begin{equation}}
\def\eeq{\end{equation}}
\def\beqn{\begin{eqnarray}}
\def\eeqn{\end{eqnarray}}


\section{Introduction}
The union of quantum field theory and general relativity has led to some of the profound predictions in modern physics. Seminal works by Hawking revealed that black holes emit thermal radiation due to quantum effects near the event horizon \cite{Hawking:1974rv, Hawking:1975vcx}. Then, Gibbons and Hawking extended this framework to cosmological spacetimes, showing that different geodesic observers experience distinct event horizons associated with a specific same temperature $T_{\rm H} \propto H$ \cite{Gibbons:1977mu}. This implies that an observer in de Sitter (dS) space perceives a thermal bath in equilibrium. While traditionally tied to the observer's static horizon \cite{Birrell:1982ix}, one possible interpretation, based on open quantum systems \cite{Alicki:2023rfv}, suggests this thermal state can be interpreted as an intrinsic property of the global vacuum \footnote{One can also refer to recent Ref.\cite{Li:2025azq} to see more intriguing results from dS thermal effects.}.

The application of Hawking radiation to produce primordial gravitons was first explored with light, evaporating primordial black holes \cite{Anantua:2008am, Dolgov:2011cq, Dong:2015yjs, Hooper:2020evu, Arbey:2021ysg, Cheek:2022dbx, Ireland:2023avg}. Yet, radiation from the cosmological event horizon has received comparatively little attention. Although some early work has referred to radiation density forms from the dS universe \cite{Gott:1982zf, Padmanabhan:2002ji, Hu:2010tx, Barrau:2014kza}, it has not considered the value of the graviton component of radiation for probing the thermal history of the early universe. Cosmic inflation, namely a phase of quasi-dS expansion in the early universe \cite{Guth:1980zm, Linde:1981mu, Albrecht:1982wi}, provides a natural lab for these quantum effects. The standard theory of primordial Gravitational Waves (GWs) focuses on vacuum fluctuations that are stretched by the rapid expansion. As these modes cross the Hubble horizon $k<aH$, their amplitudes freeze, resulting in a scale-invariant stochastic background that reenters the horizon later \cite{Starobinsky:1979ty}. These frozen super-horizon modes are the primary target of current GW searches, e.g., CMB B-mode polarization and interferometers \cite{Caprini:2018mtu, Christensen:2018iqi, Kolb:2023ydq}.

However, distinct from these frozen super-horizon fluctuations, the sub-horizon modes $k>aH$ inside the dS horizon remain in thermal equilibrium. Instead of freezing, these modes behave as a local thermal reservoir, as Alicki et al.\cite{Alicki:2023rfv} proved, using quantum thermodynamics and open quantum systems, that a local system (detector) placed in dS space would thermalize. It provides a possibility that while the geometric definition of a static horizon is observer-dependent, the physical thermal state is intrinsic to the vacuum and ubiquitous. Due to the homogeneity of the inflationary spacetime, this thermal state is locally realized at every point in the universe. This implies that the background field actively transfers energy to the detector, distinguishing the thermal bath from a mere coordinate artifact. Consequently, this energy density represents a physical reservoir that is subject to thermodynamic conservation laws; it cannot arguably disappear upon the cessation of the horizon but may be released into the bulk. A natural question arises: what happens to these thermal gravitons when inflation ends? 

From a perspective of thermodynamic phenomenology, considering that typical inflation should end violently and non-adiabatically to transfer energy from the inflaton sector to the radiation sector, this thermal bath of gravitons is expected to be released rather than adiabatically vanishing based on thermodynamic conservation. Unlike the continuous particle production often discussed in the context of pre/reheating, we treat this signal as a thermodynamic snapshot of the thermal dS state that decouples from the horizon geometry at the end of inflation. While neglecting the microscopic details of mode evolution, this approach captures the macroscopic energy budget determined by the horizon temperature. These gravitons, having never crossed the horizon to be frozen, propagate as free particles undergoing cosmological evolution\footnote{Note that the graviton is free from any interaction at the beginning of reheating, as the decoupling temperature is almost Planck temperature.}.

This mechanism predicts a unique relic graviton background that differs significantly from the standard scale-invariant spectrum. Since the source is a thermal bath at the beginning of reheating, the resulting signal manifests as a High-Frequency (HF) Planckian spectrum. The peak frequency is sensitively determined by the reheating temperature and lies in the MHz regime for typical inflation. This frequency band is a prime target for emerging HF GW technologies \cite{Aggarwal:2025noe}. In conclusion, we calculate the present-day energy density and spectral shape of this thermal remnant, providing a new promising theoretical target for HF GW astronomy.

This work is organized as follows: Sec.II reviews the Gibbons-Hawking radiation spectrum. In Sec.III, we track its evolution and give an analytical result. We take some typical values to test these signals in Sec.IV. We end this work by concluding our results and discussing the future promise. Hereinafter, we keep the constants $\hbar= c $ = $k_{B}=1$, reverting to SI Units if necessary.

\section{Gibbons-Hawking radiation spectrum: An Open System Perspective from Quantum Thermodynamics}
We start modeling the inflationary phase to quasi-dS as a flat dS universe, with the FLRW form line element given by \cite{deSitter:1917zz}
\begin{equation}
    ds^2 = - dt^2 + a(t)^2 (dr^2 + r^2 d\Omega^2_2),
\end{equation}
and the Hubble rate is $ H_{\Lambda} = \dot{a}(t)/a(t) = \sqrt{8 \pi G \rho_\phi / 3}$, with  $\rho_\phi$  representing the energy density during the dS phase, determined by the form of the potential function of the inflation model. Here we take the \textit{Planck}2018+BK15 constraint, leads to an upper bound on the Hubble parameter during inflation of $H_{\Lambda} \lesssim 2.5 \times 10^{-5} M_{\rm Pl} \,\, {\rm ( 95\% \, CL )}$ \cite{Planck:2018jri}. Such an energy scale, though substantial, remains well below the Planck scale, avoiding significant effects of quantum gravity. The scale factor satisfies  $a(t) = H_{\Lambda}^{-1} e^{H_{\Lambda} t}$, and the cosmological event horizon radius can then be expressed as $r_{\rm H} \equiv a(t) \int_{t}^{t_{\rm max}} \frac{ d t'}{a(t')} =  H_{\Lambda}^{-1}$.

The pioneering work of Gibbons and Hawking, based on quantum field theory, derived the existence of thermal radiation in dS space with temperature \cite{Gibbons:1977mu}
\begin{equation}\label{gh}
        T_{\rm H} = \frac{ H_{\Lambda}}{2\pi} ,
\end{equation}
by identifying the horizon in static coordinates. However, the traditional derivation relies on vacuum selection and particle definitions in specific coordinates, and the universality of its physical picture is often debated. 

The starting point of Ref.\cite{Alicki:2023rfv} is to describe an open quantum system embedded in an expanding three-dimensional space $\mathbf{x}$ and evolving with cosmic time $t$. The key step is incorporating the effect of cosmic expansion into the effective dynamics of the system. Under the approximation of neglecting the backreaction of the system on spacetime, which is consistent with standard treatments in quantum field theory in curved spacetime, the Hamiltonian of the system can be written as
\begin{equation}
        \hat{H}_D(t) = \hat{H} + h(t) \hat{D}.
\end{equation}
Here, $\hat{H}$ is the system Hamiltonian in static space, $h(t)$ is the Hubble parameter, and 
$\hat{D} = \frac{1}{2}(\hat{\mathbf{x}} \cdot \hat{\mathbf{p}} + \hat{\mathbf{p}} \cdot \hat{\mathbf{x}})$ is the spatial dilation generator, namely compression operator. This decomposition allows us to place the entire \textit{system + background field} within the standard framework of open quantum system theory for analysis.

Consider a simple localized detector weakly coupled to a quantum scalar field in the dS background. In open system theory, the influence of the environment (here, the background field) on the system is completely characterized by the spectral density function $\tilde{G}(\omega)$, which is the Fourier transform of the environmental two-point correlation function $G(t) = \langle \hat{\phi}(t) \hat{\phi}(0) \rangle$.

The crucial achievement of Ref.\cite{Alicki:2023rfv} is the direct calculation of this spectral density in the dS vacuum state $|\Omega \rangle$, satisfying the Bunch-Davies condition. By solving the Heisenberg equation for the field operator evolving under the Hamiltonian $\hat{H}_D$ (taking $h = \text{const.}$) and introducing an UV regularization cutoff $\Lambda$, they obtained the unregulated spectral density expression
\begin{equation}
        \tilde{G}^{\text{dS}}_{\Lambda}(\omega) =  \frac{ 2\pi^2\Lambda^2 \omega}{h^2+\Lambda^2} \operatorname{csch}\!\left(\frac{\pi\omega}{h}\right) \exp\left[ \frac{2\omega}{h} \arcsin\left(\frac{\Lambda}{\sqrt{\Lambda^2 + h^2}}\right) \right]
\end{equation}
Analysis of this expression reveals a crucial fact: it satisfies the Kubo-Martin-Schwinger (KMS) condition, a hallmark of thermal equilibrium. Specifically, the relation $\tilde{G}_{\Lambda}(-\omega) / \tilde{G}_{\Lambda}(\omega) = e^{-\beta(\Lambda) \omega}$ defines an \textit{effective} inverse temperature related to the cutoff $\Lambda$ as
\begin{equation}
        \beta(\Lambda) = \frac{4}{h} \arcsin\left(\frac{\Lambda}{\sqrt{\Lambda^2 + h^2}}\right)
\end{equation}
Physically observable temperature should be universal after removing microscopic details, i.e., taking the limit $\Lambda \to \infty$. In this limit, the effective temperature \textit{condenses} into a cutoff-independent constant
\begin{equation}
        \lim_{\Lambda \to \infty} \beta(\Lambda) = \frac{2\pi}{h} \quad \Rightarrow \quad T_{\text{dS}} \equiv \frac{1}{\beta} = \frac{h}{2\pi}
\end{equation}
One can find that we recover the Gibbons-Hawking temperature as \eqref{gh} if we take the Hubble parameter $h=H_\Lambda$.

Simultaneously, the spectral density function simplifies to a result with a clear thermodynamic form
\begin{equation}\label{G}
        \tilde{G}^{\text{dS}}(\omega) \equiv \lim_{\Lambda \to \infty} \tilde{G}^{\text{dS}}_{\Lambda}(\omega) = \frac{(2\pi)^2 \omega}{1 - e^{-\omega / T_{\text{dS}}}}.
\end{equation}
This expression has the typical structure of a Planck distribution, where $\omega / (1 - e^{-\omega/T})$ is characteristic of the spectral density of a bosonic thermal bath.

In open system theory, the spectral density $\tilde{G}(\omega)$ completely characterizes the system-environment coupling. To obtain the energy density of the thermal bath itself, a macroscopic quantity necessary for calculating the relic GW background, we need to derive the thermodynamic properties of the system from the spectral density. For a general thermal bath, the relation between spectral density and density of states $n(\omega)$ is
\begin{equation}
        \tilde{G}(\omega) = \frac{8\pi^4 n(|\omega|)}{\omega(1-e^{-\beta\omega})}.
\end{equation}
Substituting the dS spectral density \eqref{G} into the above, we obtain the density of states in dS space
\begin{equation}
        \frac{8\pi^4 n^{\text{dS}}(\omega)}{\omega(1-e^{-\omega/T_{\text{dS}}})} = \frac{(2\pi)^2 \omega}{1-e^{-\omega/T_{\text{dS}}}} \quad \Rightarrow \quad n^{\text{dS}}(\omega) = \frac{\omega^2}{2\pi^2}.
\end{equation}
This is precisely the standard density of states for massless particles in three-dimensional space. 

One might question an apparent \textit{scale mismatch}: the temperature $T_{\rm ds}$ is set by the Hubble scale, yet modes with $k \gg ah$ carry this thermal distribution. This is resolved by noting that the Bunch-Davies vacuum becomes thermal at \(T_{\rm dS}\) upon tracing out the sub-horizon modes. Concretely, the open-quantum-system approach to stochastic inflation yields a Lindblad equation with noise coefficient $h^3/(4\pi^2)$, satisfying the KMS condition and forcing a Bose–Einstein occupation $n(\omega)=1/(e^{\omega/T_{\rm dS}}-1)$ \cite{Li:2025azq}. Operationally, an environment that thermalizes a subsystem via a Planck spectrum and carries macroscopic energy density is indistinguishable from a genuine heat bath.

Consequently, for a massless bosonic thermal bath at temperature $T_{\text{dS}}$, the spectral energy density per unit volume is \cite{Alicki:2023rfv}
\begin{equation}
        \frac{d\rho}{d\omega} = g_i \frac{n^{\text{ds}}(\omega) \omega}{e^{\omega/T_{\text{dS}}} - 1} = \frac{g_i}{2\pi^2} \cdot \frac{\omega^3}{e^{\omega/T_{\text{dS}}} - 1},
\end{equation}
where $g_i$ is the number of bosonic freedom degrees. For gravitons with spin-2, two polarizations, the spectral energy density per unit volume should be written as\footnote{Gravitons, as $s=2$ bosons, are a common and reasonable approach in calculating the energy density of GWs, and adopted by many studies on black hole radiation \cite{Dolgov:2000ht, Anantua:2008am, Dolgov:2011cq, Dong:2015yjs, Hooper:2020evu, Arbey:2021ysg, Cheek:2022dbx, Ireland:2023avg}. However, to maintain rigor, we refer to it as the graviton background rather than specifically attributing it to the emission of individual gravitons. Although the language of particles is employed for intuition, our predictions are formulated entirely in terms of gauge-invariant energy densities and spectra, which are directly related to observable GW backgrounds.}
\begin{equation}\label{rho}
        \frac{d\rho_{\text{G}}}{d\omega} = \frac{1}{\pi^2} \frac{\omega^3}{e^{\omega / T_{\text{H}}} - 1}.
\end{equation}
by writing the temperature as $T_{\text{H}} \equiv T_{\text{dS}}$. Now one can write the total power of graviton radiation integrated over all frequencies $\omega$, and find it obeys the Stefan-Boltzmann law as \cite{Padmanabhan:2002ji, Alicki:2023rfv}
\begin{equation}\label{r}
        \rho_{\text{G}} = \int_0^\infty \frac{d\rho_{\text{G}}}{d\omega} d\omega = a_d \, T_{\text{H}}^4.
\end{equation}
where $a_d = \pi^2/15 $ is the radiation density constant. This result has been consistently recovered in various cosmological contexts, from the open universe models of Gott \cite{Gott:1982zf} to the bouncing cosmologies of Barrau et al.\cite{Barrau:2014kza}. But unlike earlier works that focus on the existence of a steady energy density associated with dS horizons, the signal studied here should be viewed as a snapshot of the thermal state at the moment inflation ends.

It is important to address a common theoretical concern regarding such a steady thermal state, namely the Trans-Planckian Problem. In a standard particle flux interpretation, maintaining a constant energy density during expansion would require a continuous inflow of modes with sub-Planckian wavelengths. However, within the framework of Alicki et al.\cite{Alicki:2023rfv}, the energy density in Eq.\eqref{r} is not a collection of redshifting particles, but an intrinsic thermodynamic potential of the vacuum field interacting with the background geometry. The energy is dynamically maintained by the work done by the gravitational field (the metric expansion) on the quantum field, similar to how the cosmological constant maintains constant density. Therefore, the constant $\rho_{\rm G}$ does not require replenishing by a particle flux from a trans-Planckian reservoir, but is a global stationary state of the vacuum.

We emphasize that this mechanism operates strictly within the standard cold inflation framework and must not be confused with warm inflation. 
In warm inflation, a genuine radiation fluid is sustained by dissipative particle production from the inflaton, which modifies the background dynamics and suppresses the tensor-to-scalar ratio, restricting the Hubble scale to \(H\ll 10^{12}\,\mathrm{GeV}\).
Our scenario, by contrast, involves no direct coupling between the graviton and the inflaton; the background evolution and curvature perturbations are completely unaffected, and the Hubble parameter retains its full cold-inflation range.
The thermal environment studied here is not a classical radiation fluid but an intrinsic statistical property of the Bunch–Davies vacuum. 
This is explicitly realized in the open-quantum-system formulation of stochastic inflation \cite{Li:2025azq}: tracing over sub-horizon modes yields a Lindblad equation with a stationary diffusion coefficient that satisfies the KMS condition at $T_{\rm dS}$. The noise kernel is sustained by the background expansion during the quasi-dS phase, does not suffer from $a^{-4}$ dilution, and requires no ad hoc particle production.

One may notice that \eqref{r} is a perfect spectrum. We parameterize the non-adiabatic release of the thermal graviton bath at the end of inflation by an efficiency factor $\gamma \leq 1$, which encodes the microscopic details of the transition that are beyond the scope of the present thermodynamic treatment. In the typical framework of inflation, the release efficiency factor\footnote{The factor $\gamma$ characterizes the conversion of the dS thermal bath into freely propagating radiation at the moment of horizon disappearance, not the subsequent reheating evolution.} $\gamma \approx1$ for the local thermal graviton background at the end of inflation is a plausible expectation based on three core theoretical supports:
\begin{itemize}
    \item The Rapid End of Inflation: The inflation field will roll down to the potential minimum within an extremely short timescale, much less than a Hubble time $\Delta t \ll H^{-1}$. And oscillation time $m_\phi^{-1}  \lesssim H^{-1}$ is almost instantaneous. This rapid change, almost faster than the Hubble time, implies that the background geometry acts on the thermal graviton modes as a \textit{quantum quench}. The sub-horizon thermal graviton modes do not have sufficient time to adiabatically adjust their wavefunctions to the new background\footnote{The release occurs at $\Delta t_{\rm end}$, while the reheating evolution takes place at $\Delta t_{\rm reh}\gg \Delta t_{\rm end}$. It is crucial to distinguish between the non-adiabatic termination of the dS phase and the subsequent reheating dynamics. This process is independent of whether reheating proceeds instantaneously or gradually. }. Their statistical distribution at the end of inflation is frozen and serves as the initial condition for subsequent free evolution.
    \item The Extreme Flatness Eliminating Geometric Scattering: After a sufficient $N > 50$ of e-folds of expansion, the spatial curvature parameter $|\Omega_k| \ll 1$. This directly implies that in the local spacetime region at the end of inflation, the geometry can be described with approximation by a Minkowski spacetime. Graviton wave equations lack coupling terms to background curvature. 
    \item The Minuscule Scattering Cross-section of Gravitons: Graviton interaction vertices are suppressed by inverse powers of the Planck mass as $\sigma \propto  M_{\text{Pl}}^{-4}$. Therefore, the probability that the thermal graviton background undergoes inelastic scattering with other components (including itself) in the post-inflation universe, thereby losing energy or altering its spectral shape, is entirely negligible.
\end{itemize}
Therefore, under the typical picture of inflation, there is currently no efficient energy loss mechanism known in the transition from a local thermal bath to a freely propagating relic background, leading to an order-of-magnitude expectation $\gamma\approx1$. Future observation finding a signal amplitude significantly $\gamma\ll1$ might itself become an important clue, revealing new physics beyond the typical inflation picture. Although non-equilibrium dynamical processes at the microscopic scope may lead to non-uniform factors, the time-dependent Bogoliubov transformation in the dS spacetime still faces significant difficulties at present. As a semi-analytical prediction, this result is currently the most likely option as a phenomenological approximation.

One can also find the extreme value for \eqref{rho}, which is consistent with Wien's displacement law. So the frequency associated with the peak radiation energy density at this temperature is written as
\begin{equation}
    f_{\rm peak}=\frac{x}{2\pi} T_{\rm H},
\end{equation}
where $x \approx 2.8214$ is a constant solved with the Lambert $W$ function. Here, we convert circular frequency $\omega$ to linear frequency $f = \omega /2\pi$ for convenience.

\begin{figure}[h]
    \centering
    \includegraphics[width=1\linewidth]{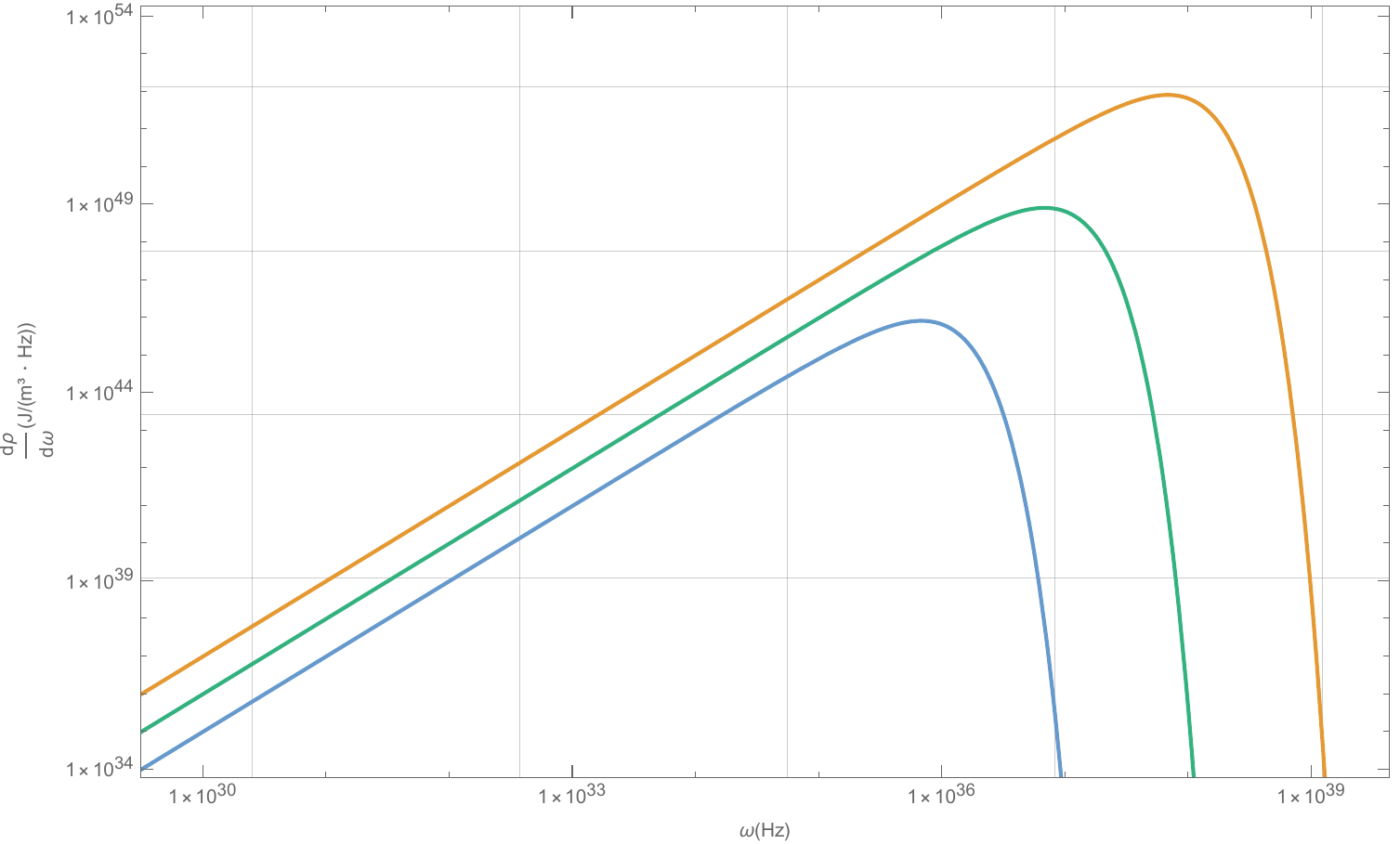}
    \caption{The plot of energy density spectra versus the frequency at the instant of reheating, shown for $\gamma = 1$. We select three sets of traditional inflationary Hubble values $H_\Lambda \sim 10^{12}$ (blue), $10^{13}$ (green), and $10^{14}\, \text{GeV}$ (orange), respectively.}
    \label{h}
\end{figure}

We plot these radiation energy density spectra in Fig.\ref{h} by selecting three sets of typical Hubble values $H_\Lambda$ during the dS period. One can easily find that these spectra share the same profile as the black hole spectra, but here the decisive influencing factors come from the $H_\Lambda$, not the black hole temperature. One can also find that the larger the $H_\Lambda$, the higher the radiation energy density can reach, which suggests that more violent inflation would result in more intense signals.

Here, once again, we emphasize that the mathematical results above have profound physical implications:
\begin{itemize}
    \item Thermalization: In open system theory, a spectral density satisfying the KMS condition as shown in \eqref{G} implies that for any localized system with a discrete energy spectrum weakly coupled to it, the unique steady state of its density matrix is the Gibbs state $\hat{\rho} \propto e^{-\hat{H}/T_{\text{H}}}$ at temperature $T_H$. The system will irreversibly relax to this thermal state from any initial state.
    \item Reality of the Thermal Bath: This thermal bath is an intrinsic physical property of the dS vacuum state, not an artifact of coordinate choice. Ref.\cite{Alicki:2023rfv} specifically notes that this temperature and thermal equilibrium property hold globally in the cosmic rest frame (comoving observer in FLRW), consistent with a real physical thermal bath \footnote{For example, the CMB radiation exhibits an ideal blackbody spectrum only in the comoving frame.}.
    \item Universality of the Energy Density: Based on the independence from spatial coordinates, it shows that the dS thermal bath possesses a real and uniform energy density $\rho_{\text{G}} \propto T_{\text{H}}^4$, providing a quantitative macroscopic description of the thermal graviton background present during inflation.
\end{itemize}

This quantum thermodynamic picture provides a clear and solid framework for understanding the graviton background during inflation. Traditional inflation theory focuses on quantum fluctuations of super-horizon modes, which are frozen. Quantum thermodynamics proves that sub-horizon modes are actually in local thermal equilibrium with the dS thermal bath, with a Planck spectrum. This bath is not a mathematical equivalence but a physical entity with effective energy density $\rho_{\text{G}}$. According to energy conservation, when the dS horizon disappears, these pre-existing thermal excitations (thermal gravitons) are unlikely to be adiabatically erased. The transition from inflation to reheating is equivalent to a \textit{quantum quench} of the background geometry. The dS thermal bath suddenly loses its heat source (the horizon geometry), and the thermal gravitons in equilibrium within it immediately decouple and are released into the universe as free, propagating radiation.

\section{Standard cosmological evolution}
In an expanding universe, the graviton energy density and frequency are not fixed but undergo cosmological redshift. Importantly, for radiative components, the total number of radiative particles is conserved, and the energy conservation equation for radiation follows $\text{d}\rho_r = -4 \rho_r \text{d}\ln a $. So these components evolve as
\begin{equation}
\rho_r \sim a^{-4}, \quad \quad f \sim a^{-1},\label{12}
\end{equation}
respectively. Denoting the end of inflation by “$*$”, we have $\rho_r a^4 = \rho_r^* a_*^4$, $f a = f_* a_*$. Consequently, the energy density at a certain time is related to the values at the end of inflation as
\begin{equation}
    \rho_{\rm G} = \rho_{\rm G}^*\left( \frac{a_*}{a} \right)^4,\,\,
    f = f^*\left( \frac{a_*}{a} \right).
\end{equation}
Note that the graviton component possess Planck scale decoupling temperature, which means it is free from any interaction, so we can just care about the scale factor determined by the temperature of different universe periods. We adopt a hypothesis that the thermal graviton background is established by the dS horizon and is fully formed by the end of the inflationary phase. The universe rapidly transitions to the decelerated Friedmann stage, where the associated Gibbons-Hawking temperature (linked to the Hubble rate) is insufficient to sustain a thermal bath of comparable energy density \cite{Brandenberger:1982xi}. 

We consider a standard picture in which reheating follows inflation, then transitions to a radiation-dominated universe. Note that readers should distinguish between the instantaneous release of radiation at the end of inflation and non-instantaneous evolution during the reheating phase. Even if reheating proceeds over many Hubble times, the thermal graviton modes have already decoupled as free radiation immediately after the geometric quench, and therefore cannot re-thermalize or adiabatically adjust their distribution. 

The temperature of reheating depends on specific inflation models, including but not limited to $T_{\rm reh, l} \sim 10^9 \, {\rm GeV}$ for $R^2$ inflation \cite{Bezrukov:2011gp}, a higher reheating temperature $T_{\rm reh, h} \sim 10^{13} \, {\rm GeV}$, as predicted in Higgs inflation \cite{Bezrukov:2008ut, Rubio:2018ogq}. The dynamics of this phase are still dominated by scalar fields, and the effective equation of state is $w \approx 0$. Therefore, in the reheating phase dominated by scalar fields, the evolution of energy density with the scale factor follows $\rho \propto a^{-3}$. So we have the relational expression of scale factor between the end of inflation and the end of reheating as $\frac{a_{\rm reh}}{a_*}=(\frac{\rho_{*}}{\rho_{\rm reh}})^{1/3}$ where $\rho_*=3 H_{\Lambda}^2 m_p^2$ ($m_p$ is reduced Planck mass) and $\rho_{\rm reh} \sim T_{\rm reh}^4$, the energy density of GWs at the end of reheating can be expressed as
\begin{equation}
   \rho_{\rm G}^{\rm reh} = \rho_{\rm G}^* \left( \frac{a_{*}}{a_{\rm reh}} \right)^4 .
\end{equation}

To examine the current energy density in the form of GWs, we note that the total energy of this radiation remains conserved throughout cosmic evolution after reheating. Consequently, the present-day energy density in the form of GWs is
\begin{equation}
   \rho_{\rm G, 0} = \rho_{\rm G}^{\rm reh} \left( \frac{a_{\rm reh}}{a_0} \right)^4 ,
\end{equation}
where $a_0 = a(t_0)=1$ is the scale factor today. Given that the preceding derivations rely on a zeroth-order approximation in the dS model, it is crucial to calibrate them against present-day observations, particularly the scale factor at the end of inflation. To estimate the scale factor at the end of inflation, we apply entropy conservation. Expressing the scale factor ratio in terms of plasma temperature and the effective entropy degrees of freedom, we have $g_{\star,s} a^3 T^3 = \text{constant}$, yielding \cite{Mukhanov:2005sc}
\begin{equation}
a_{\rm reh} = \left( \frac{g_{\star,s}(T_0)}{g_{\star,s}(T_{\rm reh})} \right)^{1/3} \frac{T_0}{T_{\rm reh}},
\end{equation}
where  $T_0 = 0.23 \, \text{meV}$  represents today's temperature of CMB and $g_{\star,s}(T_0) = 3.91$ is the effective degrees of freedom in entropy now. At the end of inflation, the universe's temperature remains exceedingly high, and nearly all standard model particle degrees of freedom are active, allowing us to approximate $g_{\star,s}(T_{\rm reh}) = 106.75$ \cite{Husdal:2016haj}.

Here, we focus on the current energy density originating from cosmological horizon radiation, characterized by the dimensionless density parameter $\Omega_{\rm G}$ at the current frequency $f_{\rm 0}$, defined by
\begin{equation}
    \Omega_{\rm G} = \frac{1}{c^2 \,\rho_{\rm crit} } \frac{d \rho_{\rm G,0}}{d \text{ln}\,f_{\rm 0}} ,
\end{equation}
where $\rho_{\rm crit} = \frac{3 H_0^2}{8 \pi G}$ represents the current critical energy density, with $H_0 = 100 h \, \text{km} \cdot \text{s}^{-1} \cdot \text{Mpc}^{-1}$, $h \approx 0.7$ being the present-day Hubble rate \cite{Planck:2018vyg, Riess:2021jrx}. Refer to (\ref{12}), one can write the evolution of the energy density versus logarithmic frequency as
\begin{equation}
     \frac{d \rho_{\rm G,0}}{d \text{ln}\,f_{\rm 0}}= \frac{d \rho_{\rm G}}{d \text{ln}\,f}\left( \frac{a}{a_0} \right)^4,
\end{equation}

Given the expression for today's energy density and using the relation $\frac{1}{d \text{ln}\,f}=f \frac{d}{d f}$, we finally arrive at the prediction for the dimensionless density parameter today
\begin{equation}\label{GW}
{\begin{aligned}
    \Omega_{\rm G} &=  \frac{f}{c^2 \,\rho_{\rm crit} } \frac{d \rho_{\rm G}}{d \text{ln}\,f}   \left( \frac{a_{*}}{a_{\rm reh}} \right)^4 \left[\left( \frac{g_{\star,s}(T_0)}{g_{\star,s}(T_{\rm reh})} \right)^{1/3} \frac{T_0}{T_{\rm reh}}\right]^4\\
    &\simeq \frac{16 \pi^2 \hbar  }{c^5 \,\rho_{\rm crit} }\frac{f^4 }{e^{\frac{2\pi \hbar f }{k_{B} T_{\rm H}}} - 1}   \left( \frac{a_{*}}{a_{\rm reh}} \right)^4 \left[\left( \frac{g_{\star,s}(T_0)}{g_{\star,s}(T_{\rm reh})} \right)^{1/3} \frac{T_0}{T_{\rm reh}}\right]^4
    \end{aligned}}
\end{equation}
with a current frequency, which also experienced the redshift, written as
\begin{equation}\label{f}
    f_{0} = f   \left( \frac{a_{*}}{a_{\rm reh}} \right) \left[\left( \frac{g_{\star,s}(T_0)}{g_{\star,s}(T_{\rm reh})} \right)^{1/3} \frac{T_0}{T_{\rm reh}}\right].
\end{equation}
Finally, one can find that the dimensionless density parameter today can be written a more concise form as
\begin{equation}
\Omega_{\rm G} (f_0)\simeq \frac{16 \pi^2 \,\hbar\,  }{c^5 \,\rho_{\rm crit} } \, f_{0}^4 \, F(f_{0})
\end{equation}
with a Planck-like distribution function
\begin{equation}
 \resizebox{.99\hsize}{!}{${F(f_{0})= \left(\text{exp}\left({\frac{2\pi \hbar f_0 }{k_{B} T_{\rm H}}  \left( \frac{a_{*}}{a_{\rm reh}} \right)^{-1} \left[\left( \frac{g_{\star,s}(T_0)}{g_{\star,s}(T_{\rm reh})} \right)^{1/3} \frac{T_0}{T_{\rm reh}}\right]^{-1}}\right)- 1\right)^{-1}}$}.
\end{equation}

\section{Concrete examples and observational prospects}
In this part, we will take some typical values to test the above scheme for generating the graviton background. There is no direct limit on the Hubble rate $H_\Lambda$ during inflation, because $H_\Lambda$ is determined by the specific inflationary potential $V(\phi)$. In this work, we do not directly restrict the specific form of the inflationary potential function but expect a more general phenomenological possibility, so we take the typical values under the constraints of various cosmological observations (especially, constraints from CMB \cite{Planck:2018jri}). So we choose $H_\Lambda \sim 10^{12}, 10^{13}$ and $10^{14}\, \text{GeV}$ as the typical Hubble rate during inflation in this work. 

The above choice of the Hubble rate $H_\Lambda$ means that we have placed some constraints on the first half of the RHS in \eqref{GW} and \eqref{f}. Now, notice the contents in square brackets, which are closely related to the energy scale at the beginning of reheating, and the temperature of reheating will greatly affect the evolution of the radiation density. We adopt the view, consistent with the standard evolution model of radiation-dominated phases, that typical reheating temperatures should be above $10^9\, \text{GeV}$ \cite{Kolb:1990vq, Planck:2018jri}. We have considered two sets of benchmark models: 
\begin{itemize}
		 \item  $T_{\rm reh,h} \sim 10^{13}\, \text{GeV}$ (efficient reheating case),
		 \item  $T_{\rm reh,l} \, \sim 10^{9}\,\,\,\,\text{GeV}$ (inefficient reheating case).
\end{itemize} 
Such temperatures may correspond to typical Higgs inflation and $R^2$ inflation, respectively,  which is reasonable and physically consistent. Readers should be aware that our model is not limited to these two specific reheating temperatures; instead, a wide range of parameters is available for choice. 

For the efficient reheating case, we plug in specific parameter values, and the prediction for the peak dimensionless density parameter today can now be written as 
\begin{enumerate}  
    \item  $\log_{10}(\Omega_{\rm G, h}^{\rm peak} h^2) \sim \mathcal{O}(-18)$, \,\,$H_\Lambda \sim 10^{14}\, \text{GeV}$,
    \item  $\log_{10}(\Omega_{\rm G, h}^{\rm peak} h^2) \sim \mathcal{O}(-22)$, \,\,$H_\Lambda \sim 10^{13}\, \text{GeV}$,
    \item  $\log_{10}(\Omega_{\rm G, h}^{\rm peak} h^2) \sim \mathcal{O}(-26)$, \,\,$H_\Lambda \sim 10^{12}\, \text{GeV}$,
\end{enumerate}  
at the corresponding peak frequency, written as
\begin{enumerate}  
    \item  $\log_{10}(f_{\rm peak,0,h}) \sim \mathcal{O}(8)$, \,\,$H_\Lambda \sim 10^{14}\, \text{GeV}$,
    \item  $\log_{10}(f_{\rm peak,0,h}) \sim \mathcal{O}(7)$, \,\,$H_\Lambda \sim 10^{13}\, \text{GeV}$,
    \item  $\log_{10}(f_{\rm peak,0,h}) \sim \mathcal{O}(6)$, \,\,$H_\Lambda \sim 10^{12}\, \text{GeV}$,
\end{enumerate} 
respectively. Similarly, we can also get the prediction for the inefficient reheating case as
\begin{enumerate}  
    \item  $\log_{10}(\Omega_{\rm G, l}^{\rm peak} h^2) \sim \mathcal{O}(-24)$, \,\,$H_\Lambda \sim 10^{14}\, \text{GeV}$,\\\\
     with $\log_{10}(f_{\rm peak,0,l}) \sim \mathcal{O}(6)$, 
    \item  $\log_{10}(\Omega_{\rm G, l}^{\rm peak} h^2) \sim \mathcal{O}(-28)$, \,\,$H_\Lambda \sim 10^{13}\, \text{GeV}$,\\\\
     with $\log_{10}(f_{\rm peak,0,l}) \sim \mathcal{O}(5)$, 
    \item  $\log_{10}(\Omega_{\rm G, l}^{\rm peak} h^2) \sim \mathcal{O}(-32)$, \,\,$H_\Lambda \sim 10^{12}\, \text{GeV}$,\\\\
     with $\log_{10}(f_{\rm peak,0,l}) \sim \mathcal{O}(4)$. 
\end{enumerate}

\begin{figure}[h]
    \centering
    \includegraphics[width=1\linewidth]{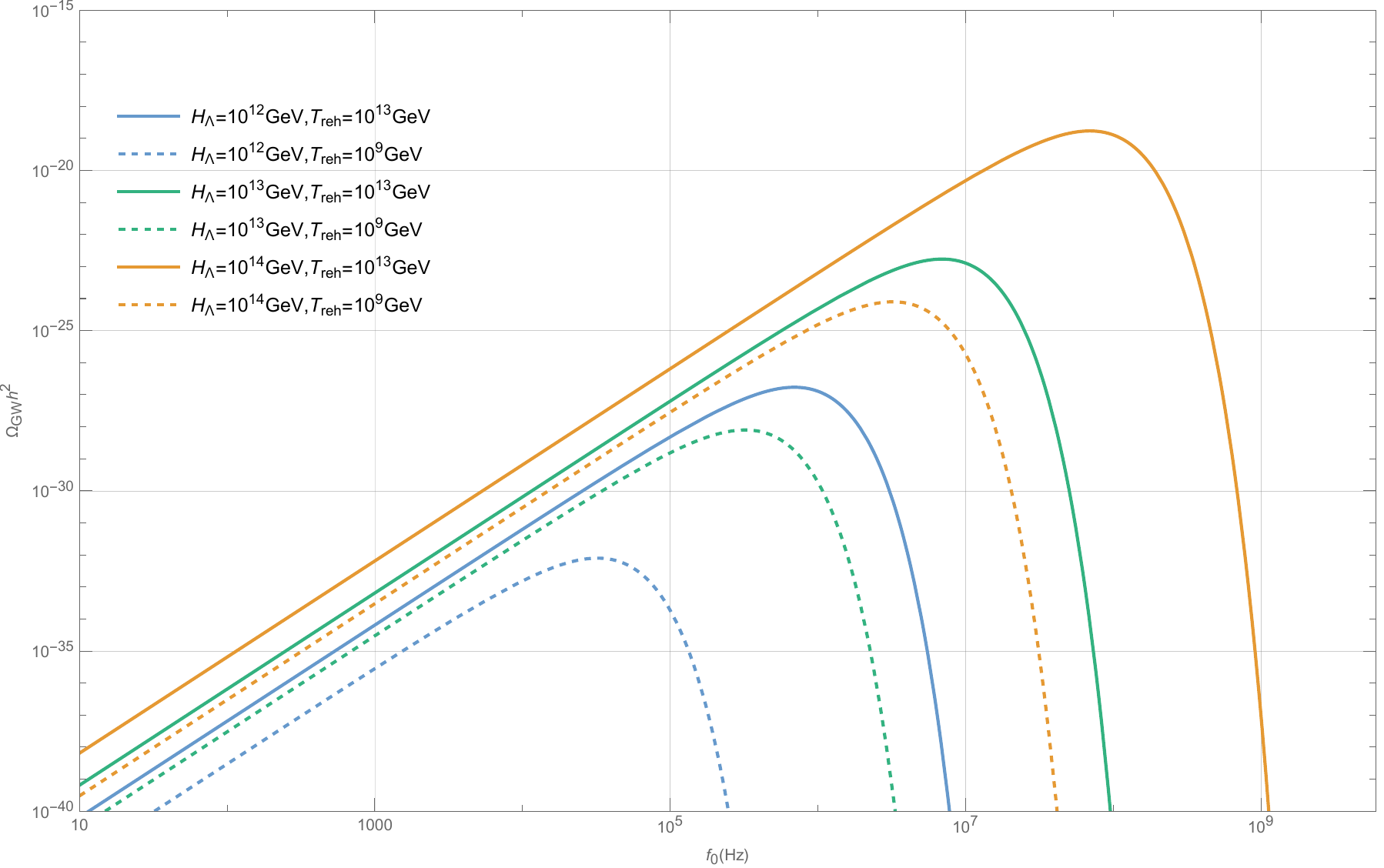}
    \caption{The dimensionless density parameter spectrum today versus frequency. We choose $H_\Lambda \sim 10^{12}$ (blue), $10^{13}$ (green), and $10^{14}\, \text{GeV}$ (orange), respectively, as the typical Hubble rate. We also set two popular reheating energy scales as $T_{\rm reh,h} \sim 10^{13}\, \text{GeV}$ (solid) and $T_{\rm reh,l} \, \sim 10^{9}\,\text{GeV}$ (dashed). This graph has been processed by log-log coordinates.
    } 
    \label{GWplot}
\end{figure}
In FIG.\ref{GWplot}, we plot our results and find that the frequency range and detection lower bound of current and planned detectors are still insufficient to cover such relic signals. We did not mark any GW sensitivity curves of detectors in Fig.\ref{GWplot}. One may refer to the literature \cite{Aggarwal:2025noe} for more details about HF GW detectors. Such signals are far from violating the tight constraints that come from CMB combined with the BBN surveys, given that the energy carried by gravitons exhibits similar characteristics to dark radiation, which contributes to the $N_{\rm eff}$ \cite{Caprini:2018mtu}, and the \textit{Planck} 2018 gives a constraint on $\Delta N_{\rm eff}< 0.3 \,\, {\rm ( 95\% \, CL )}$, leading $\Omega h^2\lesssim10^{-6}$ \cite{Planck:2018vyg}. 

However, the characteristic frequencies of these signals are not that high (MHz), and their energy density is not that low. It can be noted that the upper bound of the energy density of such relic signals is largely determined by the temperature of reheating (we take the same inflationary Hubble rate; obviously, a larger Hubble rate leads to a higher energy density). In this paper, we choose two popular reheating energy scales, and it is clear that the peak energy density (occurs at the end of the inflationary phase with condition of $H_\Lambda \sim 10^{14}\, \text{GeV}$) of the high-efficiency reheating case $\log_{10}(\Omega_{\rm G, h} h^2) \sim \mathcal{O}(-18)$ will be several orders of magnitude higher than that of the low-efficiency reheating case $\log_{10}(\Omega_{\rm G, l} h^2) \sim \mathcal{O}(-24)$, and this serves as a theoretical benchmark for the sensitivity requirements of next-generation HF GW experiments.

Then, it is worth noting that such relic signals are totally different from some well-known primordial GW signals, such as scalar-induced GWs (SIGWs) and Cosmic Gravitational Microwave Background (CGMB). The SIGWs are generated during inflation as quantum fluctuations of the metric, and it is \textit{stochastic} with an approximately scale-invariant spectrum, which means the density is approximately equal across all frequencies. However, the signals in the present work are not scale-invariant spectra, but rather distorted Planckian spectral lines. Refer to CGMB, one may find that the CGMB shares a similar spectrum with signals in Fig.\ref{GWplot}, but they are different. The CGMB is sourced by hot thermal plasma of the early Universe; however, our signals come from the effect of horizon geometry. This diversity also led to differences in the results: The peak frequency of the CGMB spectrum is
\begin{equation}
f_{\mathrm{peak}}^{\Omega_{\mathrm{CGMB}}}\approx79.8\operatorname{GHz}\left(\frac{106.75}{g_{\star s}(T_{\mathrm{max}})}\right)^{1/3} \sim 10^{11}\operatorname{Hz},
\end{equation}
with maximum estimation for the dimensionless density parameter today $\Omega_{\mathrm{CGMB}}h^2 \sim 10^{-10}$. One can find more information in FIG.4 of the Review \cite{Aggarwal:2025noe}.

Before concluding, it is imperative to mention an important study by Pi et al.\cite{Pi:2024kpw}, which meticulously evaluated the UV tail of standard primordial GWs generated by vacuum fluctuations. They demonstrated that a non-instantaneous transition from inflation to reheating produces a UV tail with a peak amplitude of approximately $\Omega_{\rm GW} \sim 10^{-16}$ (for $H_{\Lambda} \sim 10^{14}$ GeV), decaying exponentially at higher frequencies. 
Compared to this vacuum contribution, our predicted thermal remnant ($\sim 10^{-18}$ for efficient reheating) acts as a subdominant signal in a similar frequency band. The vacuum UV tail studied in \cite{Pi:2024kpw} transitions into a nearly scale-invariant spectrum on the IR side, while the thermal remnant follows a rigid Planckian distribution, meaning its IR side strongly suppresses as a Rayleigh-Jeans curve with $\Omega_G \propto f^4$. This might allow it to be isolated from a flat broadband foreground through future template matching. Furthermore, the exponential decay rate of the vacuum UV tail probes the dynamical smoothness and duration of the phase transition \cite{Pi:2024kpw}; the peak of the thermal remnant serves as a direct thermometer for the decoupling energy scale $T_{\rm reh}$. Therefore, rather than merely masking the thermal signal, the coexistence of the vacuum UV tail provides a complementary tool. Together, they outline a rich multi-component target for future HF GW observatories, demanding high-sensitivity broadband characterization.

\section{Discussion and Outlook}
Based on a framework established in the literature \cite{Alicki:2023rfv}, we propose that the sub-horizon thermal gravitons, which remain in equilibrium with the Gibbons-Hawking temperature throughout the inflationary phase, are rapidly released into the FLRW universe at the end of inflation from the perspective of thermodynamic phenomenology. The predicted signal lies in near MHz band, falling within the frequency target of emerging HF GW technologies \cite{Aggarwal:2025noe}, its characteristic amplitude poses a significant challenge for current detector concepts, requiring a substantial leap in sensitivity. 

The signal manifests as a distinct Planckian spectral shape. This distinguishes it fundamentally from the primordial GWs generated by vacuum fluctuations. While the SIGWs are nearly scale-invariant and probe the evolution of the Hubble rate $H(t)$ during inflation, the thermal remnant calculated here serves as a direct probe of the end of inflation. Specifically, the peak frequency of this remnant is sensitively dependent on the reheating temperature $T_{\rm reh}$. A detection of this HF peak would essentially act as a thermometer for the reheating epoch, providing constraints on $T_{\rm reh}$ that are complementary to those derived from CMB bounds on the spectral index $n_s$. Physically, our result can be interpreted as the thermal UV cutoff of the primordial GW spectrum. Modes with wavelengths larger than the horizon ($k<aH$) freeze out and form the scale-invariant spectrum. Modes deep inside the horizon ($k\gg aH$) never freeze but form the thermal bath described here.

We acknowledge that the microscopic scope may involve complex non-equilibrium quantum dynamics, such as time-dependent Bogoliubov transformations. While a full Bogoliubov calculation would determine the exact efficiency of this conversion, likely introducing a factor $<1$, the thermodynamic result sets the spectral shape. Future, more detailed work could investigate the impact of the specific dynamics of phase transitions on release efficiency. Our phenomenological approach provides a robust energy density benchmark based on macroscopic thermodynamic conservation, where this rapid decoupling indicates that the thermal spectrum is fixed at $T_H$, and captures the conservation of the horizon energy density at the moment of transition. As demonstrated by Alicki et al.\cite{Alicki:2023rfv}, the dS state possesses an intrinsic dynamical instability towards releasing its stored vacuum energy as on-shell radiation, effectively behaving like a superheated state ready to \textit{boil} into radiation. Crucially, our calculation assumes this thermodynamic instability is triggered at the end of the dS phase $a_*$, releasing the graviton which then decouples from the background dynamics. In Sec.III, we have explicitly accounted for the subsequent evolution of this remnant through the reheating phase, where typically $w\approx0$. This introduces a suppression factor $(a_*/a_{\rm reh})^4$, which is sourced by non-instantaneous evolution during the reheating stage. It is important to emphasize that while the overall amplitude of the signal scales linearly with the efficiency factor $\gamma$, the Planckian spectral shape and the peak frequency are determined solely by the $H_\Lambda$ at the end of inflation and the subsequent cosmological redshift. The derivation of the precise efficiency factor $\gamma$ from microscopic dynamics is a subject for future work, but this does not alter the qualitative physical picture of the thermal graviton remnant.

\section*{Acknowledgements}
We are grateful to Yen Chin Ong and Yu-Sen An for helpful discussions. This work is supported by the National Natural Science Foundation of China (NSFC) under Grant No.12175105.




\bibliography{draft.bib}

\end{document}